\documentclass[10pt,aps,pre,twocolumn,notitlepage,superscriptaddress,preprintnumbers]{revtex4-1}
\usepackage{amsmath,amssymb,amsfonts}
\usepackage{graphicx}
\usepackage{hyperref}
\usepackage{appendix}
\usepackage{nicefrac}
\usepackage{color}
\usepackage[normalem]{ulem}

\begin{document}
	\title{Heider and coevolutionary balance: From discrete to continuous phase transition}
	\author{A. Kargaran}
	\email{amir.kargaran25@gmail.com}
	\affiliation{Department of Physics, Shahid Beheshti University, G.C., Evin, Tehran 19839, Iran}
	\author{G. R. Jafari}
	\email{g\_jafari@sbu.ac.ir}
	\affiliation{Department of Physics, Shahid Beheshti University, G.C., Evin, Tehran 19839, Iran}
	\affiliation{Institute for Cognitive and Brain Sciences, Shahid Beheshti University, G.C., Evin, Tehran, 19839, Iran}
	\date{\today}
	
	\begin{abstract}
		Structural balance in social complex networks has been modeled with two types of triplet interactions. First, the interaction that only considers dynamic role for links or relationships (Heider balance), and second, the interaction that considers both individual opinions (nodes) and relationships in network dynamics (coevolutionary balance). The question is, as the temperature varies, which is a measure of the average irrationality of individuals in a society, how structural balance can be created or destroyed by each of these triplet interactions? We use statistical mechanics methods and observe through analytical calculation and numerical simulation that unlike the Heider balance triplet interaction which has a discrete phase transition, the coevolutionary balance has a continuous phase transition. The critical temperature of the presented model change with the root square of network size which is a linear dependence in thermal Heider balance.

	\end{abstract}
	\maketitle	
	\section{\label{sec:level1}Introduction} 	
	Structural balance theory is a key measure for the analysis of complex networks by investigating the impact of local interactions on the global structure \cite{heider}. This theory has been studied frequently in the context of social networks \cite{szell,altafini,samin}, international relations \cite{hart,galam,bramson}, biology \cite{Yen-Sheng,zahra,abbas} and ecology \cite{saiz}. In the analytical approach, researchers have proposed two types of local triplet interactions: (i) The interaction in which only links between individuals affect the network's temporal behavior, which is known as Heider balance or social balance, and (ii) Interaction that node and links are involved in network dynamics, which we call coevolutionary balance.
	
\begin{figure}[t]
		\centering
		\includegraphics[scale=0.78]{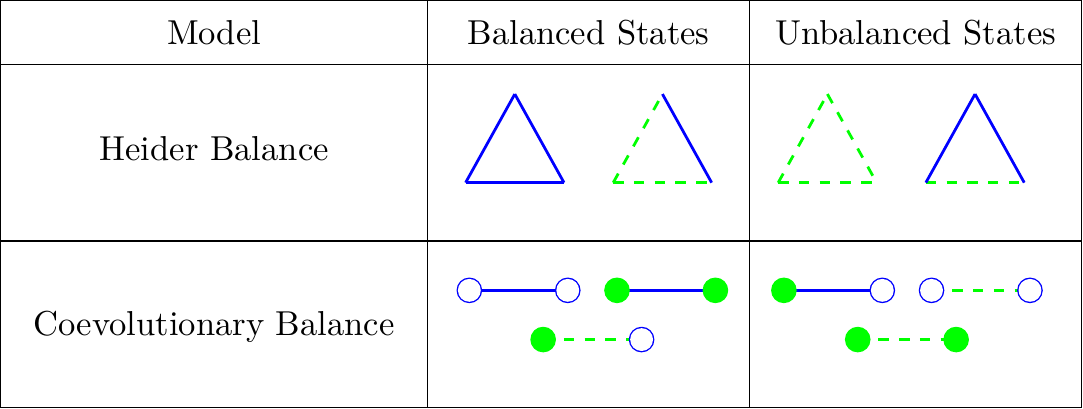}
		\caption{Balanced and unbalanced configurations of the Heider balance theory and the proposed model. Empty circles have value $ +1 $ and filled circles have $ -1 $. Solid (dashed) lines are friendly (unfriendly) links and have value  $ +1 $ ($ -1 $). }
		\label{fig:Fig1}
	\end{figure}

	The first type of triplet interaction, initially introduced by Heider \cite{heider}, just considers a role for relationships. Since its proposal, this model and its extension to graph theory\cite{cartwright}, called Heider balance, have been studied frequently in the context of social networks \cite{kirkely,hedayatifar,hassanibesheli,saeedian,sheykhali,oloomi}. This model simply considers triplet interaction between individuals (nodes) that have friendly (unfriendly) relationships (links). The axioms behind this interaction simply are ``a friend of a friend is also a friend'' and ``an enemy of my enemy is my friend'' in social networks. A triad is called balanced (unbalanced) if it contains an even (odd) number of unfriendly links (Fig.~\ref{fig:Fig1}). The unbalanced triads hold social tension and have a tendency to become balanced. The network evolves with this dynamic until no more frustration exists (heaven or bipolar) or trapped in jammed states \cite{marvel1}. In heaven state, all relationships between individuals are friendly, while in bipolar the network consists of two sub-networks with friendly (unfriendly) relations within (between) sub-networks. In jammed states, the system is trapped in a local minimum of the energy landscape.
	Modified models of Heider balance have been investigated in two scenarios: (i) Applying equilibrium statistical mechanics methods to find characteristics of equilibrium states \cite{belaza1,belaza2,masoumi,fereshteh,amir,mahsa}, (ii) Using non-equilibrium statistical mechanics and the master equation for finding stationary states properties and the time to reach them \cite{antal1,antal2,malarz,kulakowski1,kulakowski2,shojaei}.
	Furthermore, Leskovec \textit{et al.} has conducted a comprehensive comparison of this model and real social data \cite{leskovec}.
	
	The second type of local interaction takes the state of both nodes and links in a coupled dynamics, that is, the state of nodes affects the state of links and vice versa. In social science literature, this interaction consists of two parts: (i) Individuals form their beliefs based on the opinion of their neighbors, i.e., ``opinion formation''\cite{sood,castellano1}, and (ii) Network connections are organized between individuals with similar beliefs, i.e., ``homophily'' \cite{castellano2,mcpherson,bo}. Researchers have been working on a combination of the two. Holm \textit{et al.} discussed the non-equilibrium phase transition for such a coevolution model by considering a binary variable for links and a small number of opinions for nodes \cite{holm}. This phase transition depends on a single parameter that controls the balance of the two processes. Saeedian \textit{et al.} discussed such a coevolution model with considering friendly $ (+) $ and unfriendly $ (-) $, choices for links. They investigated the absorbing phase transition with coupled rate differential equations based on specific update rules \cite{meghdad1}\cite{meghdad2}. The convergence to the structural balance of triplet node-link interaction, with the aid of Hebb's principle for link adaptation, is discussed numerically by R. Singh \textit{et al.} \cite{sinha1}. They have found that the time required for this convergence has a large dispersion. Furthermore, the mean-field analysis of coupled dynamics of social balance and three-state individual opinion is investigated in a fully connected network by P. Singh \textit{et al.} \cite{sinha2}.
	
	Recently, the combination of these two types of interactions has also been investigated. G\'orski \textit{et al.} \cite{janusz} developed a framework that unifies the principles of structural balance and homophily and investigated it using non-equilibrium statistical mechanical methods. In this model, people in a society have different opinions, and the closer people's opinions are to each other, the stronger their social connection becomes (larger weight). Authors conclude that there is competition between homophily and structural balance and this competition is quite evident in societies with a limited number of opinions. Moreover, researchers have chosen the equilibrium path and, by defining an overall social tension, have investigated the equilibrium states of a social system through simulation \cite{thurner}. This study has proved that individuals' opinions in certain circumstances cause the fragmentation of society that may lead to radicalization.

	Since the properties of the first type of interaction with equilibrium and non-equilibrium methods of statistical mechanics have been well-investigated \cite{antal1, fereshteh}, the lack of investigation and comparison by equilibrium methods for the second type is noticed. In this paper, by defining an overall social tension of society (energy function which should be minimized) and considering the dynamic role for both individuals opinions and social relationships (nodes and links), we were able to investigate its analytical properties with statistical mechanics methods (Fig.~\ref{fig:Fig1}). The concept of temperature is considered as a measure of the average irrationality of individuals in society. We use \textit{exponential random graph } \cite{strauss,robins,snijders,newman1,newman2,book} as our mathematical framework and mean-field approximation for determining the averaged quantities (Sec. \ref{section two}). In  Sec. \ref{section three} the comparison between the analytical result of the proposed model and thermal Heider balance \cite{fereshteh} is presented. Moreover, the order of phase transition and the dependence of critical temperature on the network's size is discussed. Finally, we examine the analytical result via simulations (Sec. \ref{section four}).
	
	\section{Model}\label{section two}	
\begin{figure}[t]
		\centering
		\includegraphics[scale=0.9]{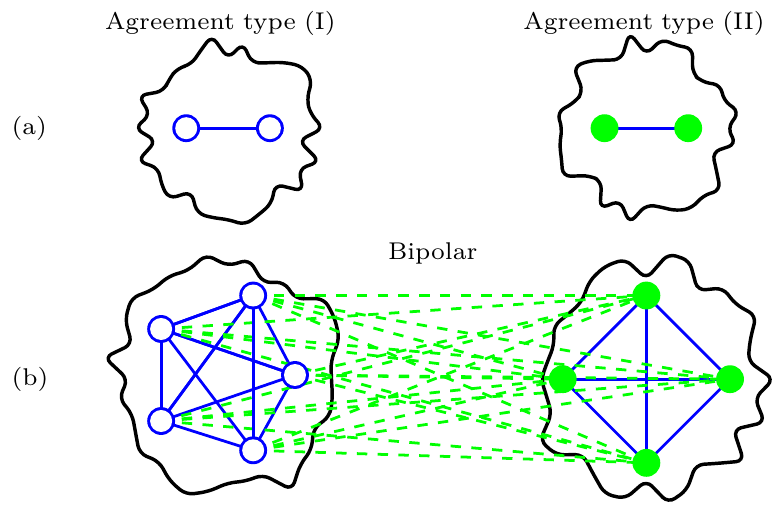}
		\caption{Balanced and unbalanced configuration of the proposed model. Empty (full) circles have the value of $ +1 $ ($ -1 $). Solid (dashed) lines are friendly (unfriendly) links and have value  $ +1 $ ($ -1 $). (a) Two kinds of consensus are observed in our model.  In each cluster, one node-link-node triplet is shown as representing the total triplets of that cluster. (b) In bipolar, all relations between the two clusters are unfriendly (disagreement). The shapes inside each cluster (square and pentagonal) refer only to the type of agreement within each cluster.}
		\label{fig:Fig2}
	\end{figure}

	According to the aforementioned, we have considered the overall social tension of society as
	\begin{equation}\label{Hamiltonian}
	\mathcal{H}(G) =-\sum_{i<j}s_{i}\,\sigma_{ij}\,s_{j},
	\end{equation}
	which should be minimized. In physics, this type of cost function is called \textit{Hamiltonian} which measures the total energy (overall social tension) of a system as a function of its configurations. $ G $ is a particular graph from all the possible configurations in a network with $ n $ individuals (nodes). Our model is based on a society that everyone knows everyone which in graph theory language, means that we have a fully connected graph. Individual $ i $  ($ j $) opinion are shown by $ s_i $ ($ s_j $), and we considered two type of opinions ($ \pm 1 $). The relationships (link) between these nodes is $ \sigma_{ij} $, which is an element of our symmetric adjacency matrix and can take friendly (unfriendly) relationships ($ \pm 1 $). Since we have two types for nodes and links the number of all configurations is $ 2^n\times 2^{\nicefrac{n(n-1)}{2}} $, where $ \nicefrac{n(n-1)}{2} $ is the total number of relationships (links) in a fully connected network. One of the configurations that reduce the social tension is two nodes with similar idea ( $ +1 $ or $ -1 $) when the link (relationships) between them is friendly ($ \sigma_{ij}=+1 $), we called this configuration the ``agreement"(consensus). Another configuration that minimized social tension as agreement is ``disagreement", in which two nodes with different opinions are linked with an unfriendly link ($ \sigma_{ij} =-1 $). We call these three configurations balanced, which they have minimal energy (social tension) and otherwise they are unbalanced [Fig.~\ref{fig:Fig1}]. The minimum overall social tension Eq. \ref{Hamiltonian} in a network can reach when all nodes and their relation are in agreement or disagreement. In these states we have two configurations for heaven state (agreement type I and II) because we have two values for nodes, and bipolar state. The illustration of the balanced states of our model is shown in Fig.~\ref{fig:Fig2}, schematically.
		
	According to the definition of overall social tension Eq. \ref{Hamiltonian} and using statistical mechanical techniques, a mathematical framework can be constructed for the analytical study of the proposed model. This mathematical framework is known as \textit{exponential random graph} \cite{strauss,robins,snijders,newman1,newman2,book} and can be used to calculate the probability of obtaining a particular configuration. This probability is written as $ \mathcal{P}(G)\propto e^{-\beta\mathcal{H}(G)} $ where $ G $ is a specific graph configuration and $ \beta $ has the role of an inverse temperature $\beta=1/T$. 
	
	The parameter $T$ is known as \textit{social temperature} \cite{thurner}, is a measure of the average volatility of individuals in society \cite{Bahr}. This means that at high temperatures, regardless of the reduction of overall social tension, people are more likely to change their opinions and relationships. At low temperatures, individuals appear more stubborn to their opinions and social ties. In other words, at high temperatures, people are more likely to behave irrationally.

	\subsection{Mean-field solution}	
	In this section we have calculated the mean quantities such as the mean of nodes $ m $ and the node-link correlation $ q $, with mean-field approximation. Suppose $ \mathcal{H}' $  is sum of all terms in Hamiltonian Eq.~\ref{Hamiltonian} that contain $ s_i $
	\begin{equation}\label{localfield}
	-\mathcal{H}_{i}=s_i \sum_{j\neq i}\sigma_{ij}\,s_j,
	\end{equation}
	and we have labeled the remaining terms as $ \mathcal{H}' $, so we have $ \mathcal{H}=\mathcal{H}'+\mathcal{H}_{i} $. We can infer from statistical mechanics
	\begin{equation}
	m\equiv\langle s_i\rangle=\sum_{G}s_i\,\mathcal{P}(G),
	\end{equation}
	where $ \mathcal{P}(G)=e^{-\beta\mathcal{H}(G)}/\mathcal{Z} $ is the Boltzmann probability and $\mathcal{Z}=\sum_{G}e^{-\beta\mathcal{H}(G)} $ is the partition function, we have
	\begin{equation}\label{single-edge}
	\begin{aligned}
	m&=\frac{1}{\mathcal{Z}}\sum_{\{s\neq s{i}\}}e^{-\beta\mathcal{H}'}\sum_{s_{i}=\pm 1}s_{i}\,e^{-\beta\mathcal{H}_{i}} \\
	&=\frac{\sum_{\{s\neq s_{i}\}}e^{-\beta\mathcal{H}'}\left[e^{-\beta\mathcal{H}_{i}(s_{i}=+1)}-e^{-\beta\mathcal{H}_{i}(s_{i}=-1)}\right]}{\sum_{\{s\neq s_{i}\}}e^{-\beta\mathcal{H}'}\left[e^{-\beta\mathcal{H}_{i}(s_{i}=+1)}+e^{-\beta\mathcal{H}_{i}(s_{i}=-1)}\right]}\\
	&=\frac{\left\langle e^{-\beta\mathcal{H}_{i}(s_{i}=+1)}-e^{-\beta\mathcal{H}_{i}(s_{i}=-1)}\right\rangle_{G'}}{\left\langle e^{-\beta\mathcal{H}_{i}(s_{i}=+1)}+e^{-\beta\mathcal{H}_{i}(s_{i}=-1)}\right\rangle_{G'}},
	\\
	\end{aligned}
	\end{equation}	
	where $ \langle \cdots\rangle_{G'} $ is obtained from the average over all graph configurations that does not contain $ s_i $. Using the similarity with spin models, it can be said that the Eq.~\ref{localfield} expresses a local field coupled to node $ s_i $. Here we have applied the mean-field approximation by substituting $\sigma_{ij}s_j\rightarrow q\equiv \langle \sigma_{ij}s_j\rangle$, we can write
	\begin{equation}\label{single-node}
	m=\frac{e^{\beta(n-1)q}-e^{-\beta(n-1)q}}{e^{\beta(n-1)q}+e^{-\beta(n-1)q}}=\tanh\left(\beta(n-1)q\right).\\
	\end{equation}	
	
	For calculating the node-link correlation $ (q) $, similar to above, we have divided the Hamiltonian sum into two sums as $ \mathcal{H}=\mathcal{H}''+\mathcal{H}_{ij}$. The $ (\mathcal{H}_{ij}) $ contains all terms that have $ s_i $ or (and) $ \sigma_{ij} $, that is
	\begin{equation}
	-\mathcal{H}_{ij}=s_i\, \sigma_{ij}\,s_j+s_j \sum_{k\neq i,j}\sigma_{jk}\,s_k,
	\end{equation}
	\begin{figure}[t]
		\centering
		\includegraphics[scale=0.21]{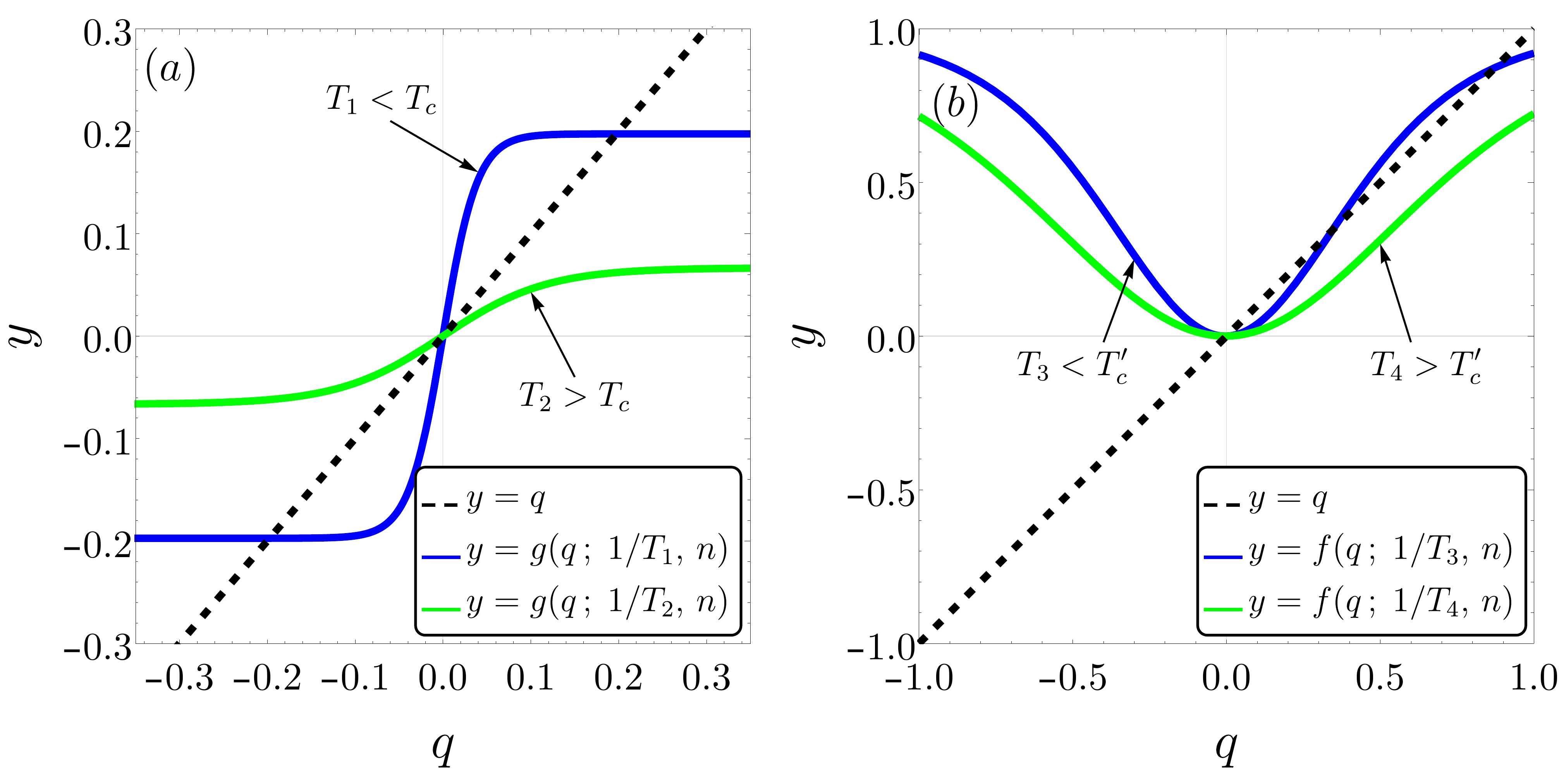}
		\caption{(Color Online) The graphical solutions for coevolutionary balance (a) and thermal Heider balance model (b) above and below their critical temperatures ($ T_c $ and $ T'_c $). The parameter $ q $ in our model is node-link correlation but in thermal Heider balance model \cite{fereshteh} is correlation between two links with a common edge. Number of nodes is 128.}
		\label{fig:Fig3}
	\end{figure}
	and let $(\mathcal{H}'')$ be the remaining terms. We can write	
	\begin{equation}\label{correlation}
	\begin{aligned}
	q&\equiv\langle \sigma_{ij}s_j\rangle=\sum_{G}\sigma_{ij}s_j\mathcal{P}
	(G)\\
	&=\frac{\sum_{G''}e^{-\beta\mathcal{H}'}\sum_{\{s_j=\pm 1, \sigma_{ij}=\pm 1\}}\sigma_{ij}s_j\,e^{-\beta\mathcal{H}_{ij}}}{\sum_{G''}e^{-\beta\mathcal{H}'}\sum_{\{s_j=\pm 1, \sigma_{ij}=\pm 1\}}e^{-\beta\mathcal{H}_{ij}}},\\
	\end{aligned}
	\end{equation}	
	where $ G'' $ is all the graph configurations that does not contain $ \sigma_{ij}$ and $ s_j $. The above statement (Eq.~\ref{correlation}) is simplified as follows
	\begin{equation}\label{meanedgenode}	q=\tanh (\beta m).
	\end{equation}
	
	By substituting (\ref{single-node}) in (\ref{meanedgenode}) we can write the self consistency equations as 
	\begin{equation}\label{self-consistance}
	q=\tanh \left[\beta \tanh\left(\beta(n-1)q\right)\right]\equiv g(q\,;\;\beta,\, n).
	\end{equation}
	In Appendix.~\ref{appendix:mean quantity}, Eq.~\ref{self-consistance} has been derived with another method. In Fig.~\ref{fig:Fig3}(a) we have illustrated a plot of $ y=q $ and $ y=g(q\,;\;\beta,\, n)$ as a function of $ q $. The intersection of two functions gives the solution of Eq.~\ref{self-consistance}. Depending on the value of temperature the functions can intersect in either one or three points in the physical region $ -1\le q\le 1 $. This behavior is the classic phenomenology of the second-order phase transition. The stability condition for solutions can be checked by the absolute value of the first derivative of right-hand side of Eq.~\ref{self-consistance} in the fixed point $ q^* $. This value is less than one for stable solutions.
	
	The critical temperature can be calculated by right-hand side of Eq.~\ref{self-consistance} in very small values of $ q $, in this limit we can use the Taylor expansion as
	\begin{equation}\label{Taylor}
	g(q\,;\;\beta,\, n) \approx q (n-1)\beta ^2+\mathcal{O}(q^3),
	\end{equation}	
	and find $T_c = \sqrt{n-1}$. The dependence of critical temperature on size is rooted in the network topology which is fully connected. In this type of network, with increasing the size, number of neighbors (in other words, the dimension) of network increases, while in famous models of physics, such as 2D Ising model, number of neighbors (dimension) of the model is fixed which results in a finite critical temperature in thermodynamics limit 
	
	Finally, we have calculated the equation for the mean of triplets $ \langle s_i\sigma_{ij}s_j\rangle $ which is the energy of network per number of triplets. By the same analogy as described in this section, we can find 
	\begin{equation}\label{energy}
	E\equiv-\langle s_i\sigma_{ij}s_j\rangle = -\tanh(\beta).
	\end{equation}
	The minus sign came from the definition of the Hamiltonian Eq.~\ref{Hamiltonian}.
	\begin{figure}[t]
		\centering
		\includegraphics[scale=0.21]{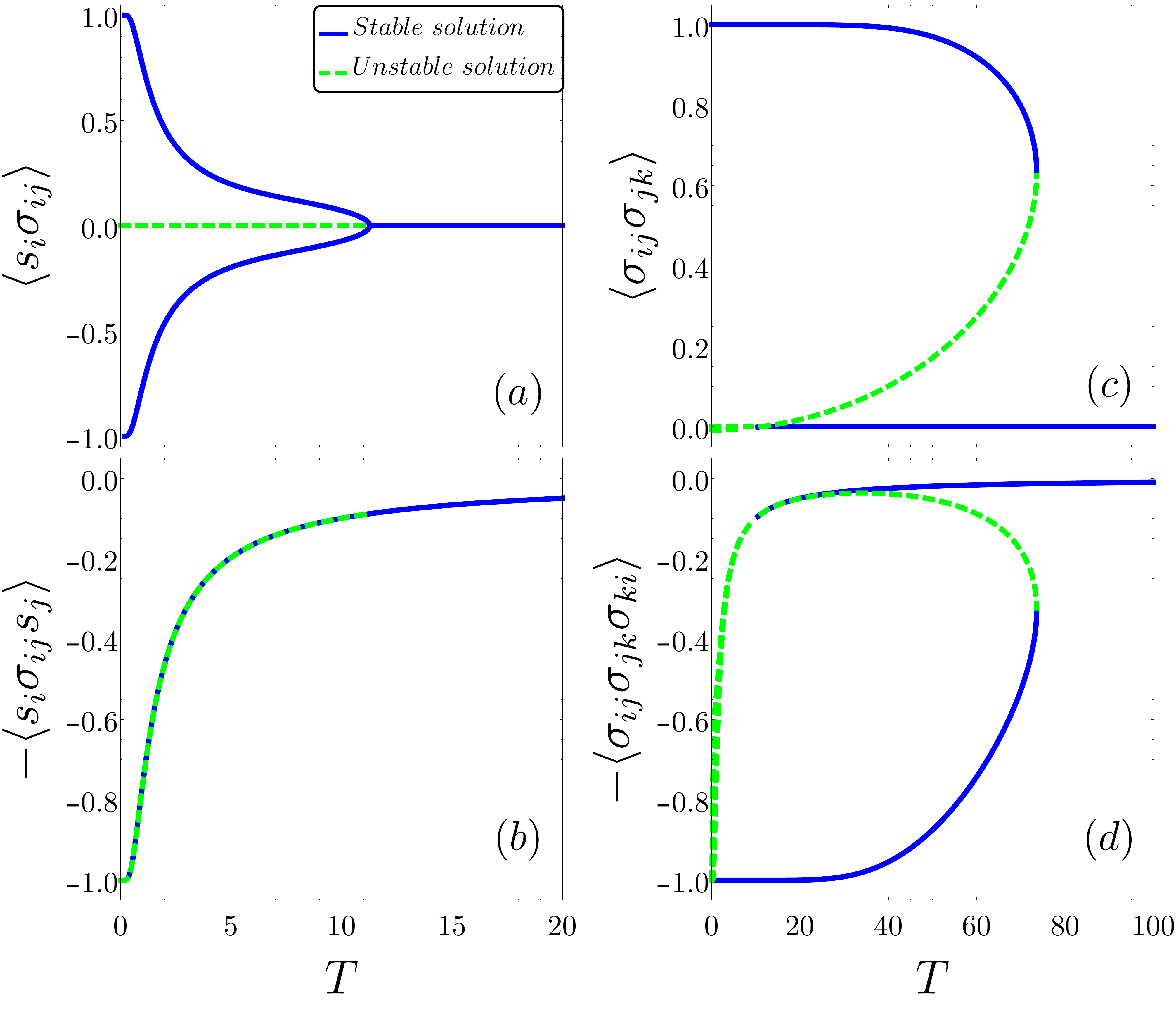}
		\caption{(Color Online) (a,b) Solution of Eq.~\ref{self-consistance} and the energy versus temperature for the presented model. (c,d) Solution of equation of state and energy versus temperature for thermal Heider balance \cite{fereshteh}. Number of nodes is 128.}
		\label{fig:Fig4}
	\end{figure}
	\section{Comparison with Thermal Heider balance }\label{section three}
	Recently, Rabbani \textit{et al.} \cite{fereshteh} have developed the statistical mechanics of the thermal Heider balance. They have considered the Heider balance Hamiltonian first introduced by Marvel \textit{et al.} \cite{marvel1} which is 
	\begin{equation}\label{fereshteh-ham}
	\mathcal{H}(G) =-\sum_{i<j<k}\sigma_{ij}\,\sigma_{jk}\,\sigma_{ki},
	\end{equation}
	where $ G $ is a specified graph and similar to our model they consider temperature as social tension. Here $ \sigma_{ij} $ is the link between nodes $ i $ and $ j $ which can take the values $ \pm 1 $ and Eq.~\ref{fereshteh-ham} is the sum of all triads in network $ G $. They have found that the phase transition between the ordered phase (heaven which means all links are positive) to the disordered phase is discrete, dependent on the temperature [Fig.~\ref{fig:Fig3}(b)]. The self-consistence equation in their model is a function of an important parameter which is $ q=\langle\sigma_{ij}\sigma_{jk}\rangle $, besides size and temperature. This parameter is the average of two stars, which is two links with a common node. They saw the hysteresis loop in their simulation which is a sign of the first-order transition.
	
	In the presented model the solutions to our equation of states (Eq.~\ref{self-consistance}) shows continuous phase transition. This equation depends on size and temperature besides an important parameter which is the node-link correlation $ q=\langle\sigma_{ij}s_j\rangle $. We have found in our simulations that the hysteresis loop does not exist in this model. In Fig.~\ref{fig:Fig4} we have compared the solutions of the self-consistence equation and energy in both models. As it can be seen in Fig.~\ref{fig:Fig4}(a), the parameter $ \langle\sigma_{ij}s_j\rangle $ for the presented model behave smoothly from one (or minus one) to zero, however in thermal Heider balance model Fig.~\ref{fig:Fig4}(c) the parameter $ \langle\sigma_{ij}\sigma_{jk}\rangle $ behaves abruptly.	
	\begin{figure}[t]
		\includegraphics[scale=0.39]{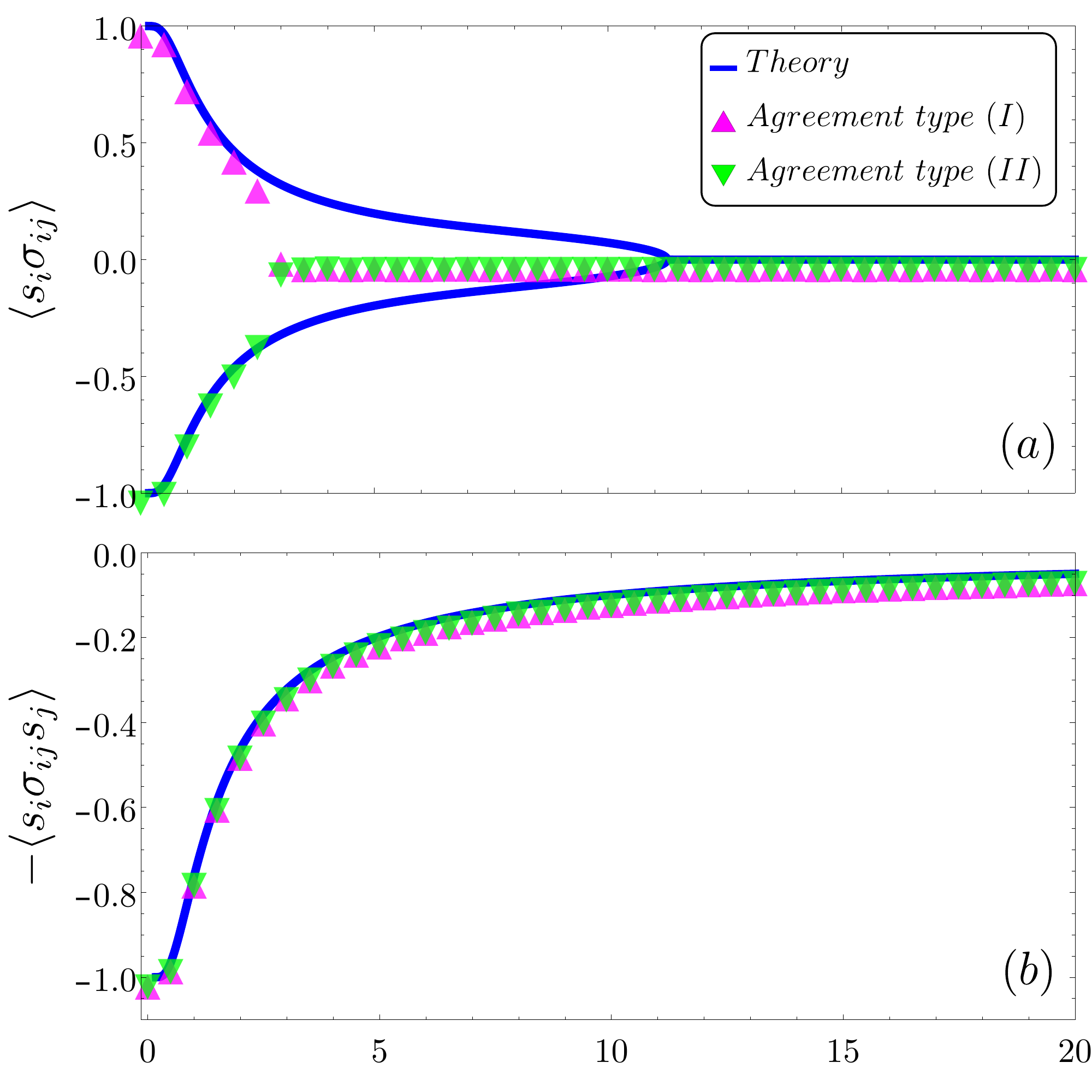}
		\caption{(Color Online) Comparison of our theory with Monte-Carlo simulations (triangles). The initial states for our simulation are agreement types I and II. (a) Mean of node-link correlation ($\langle \sigma_{ij}s_j\rangle $) versus temperature. (b) Mean of energy ($\langle s_i\sigma_{ij}s_j\rangle $) versus temperature.}
		\label{fig:Fig5}
	\end{figure}
	
	\section{Simulations}\label{section four}
	We have simulated a fully connected network in which the number of nodes ($ n $) is known. We used the Metropolis algorithm for thermalizing our network with a given temperature. In this method, we have picked a random object, which could be a node or a link, and have computed the energy for a new configuration where it is flipped. If the energy of the new configuration has decreased with respect to the old configuration the flip will be accepted otherwise if the energy increased the acceptance ratio is the Boltzmann probability. In each step of iteration, we have picked just one object: A node with probability $ p=\nicefrac{n}{(n+n(n-1)/2)} $ and a link with $ 1-p $. This probability ($ p $) is the number of nodes over the number of updating objects which is the number of nodes plus the number of links. The parameter is useful for updating all objects uniformly. In Fig.~\ref{fig:Fig5} we have illustrated our theory with Monte-Carlo simulations.
	
	In Fig.~\ref{fig:Fig5} (a) we have compared our simulation (for all initial conditions either agreement type I and II as in Fig.~\ref{fig:Fig2}) with theory. For node-link correlation ($ q $), the simulation and theory are in good agreement in low temperatures but the mean of energy is in good agreement with simulation for all range of temperatures [Fig.~\ref{fig:Fig5} (b)]. This agreement holds for the random initial condition as well. 
	
\begin{figure}[t]
		\includegraphics[scale=0.29]{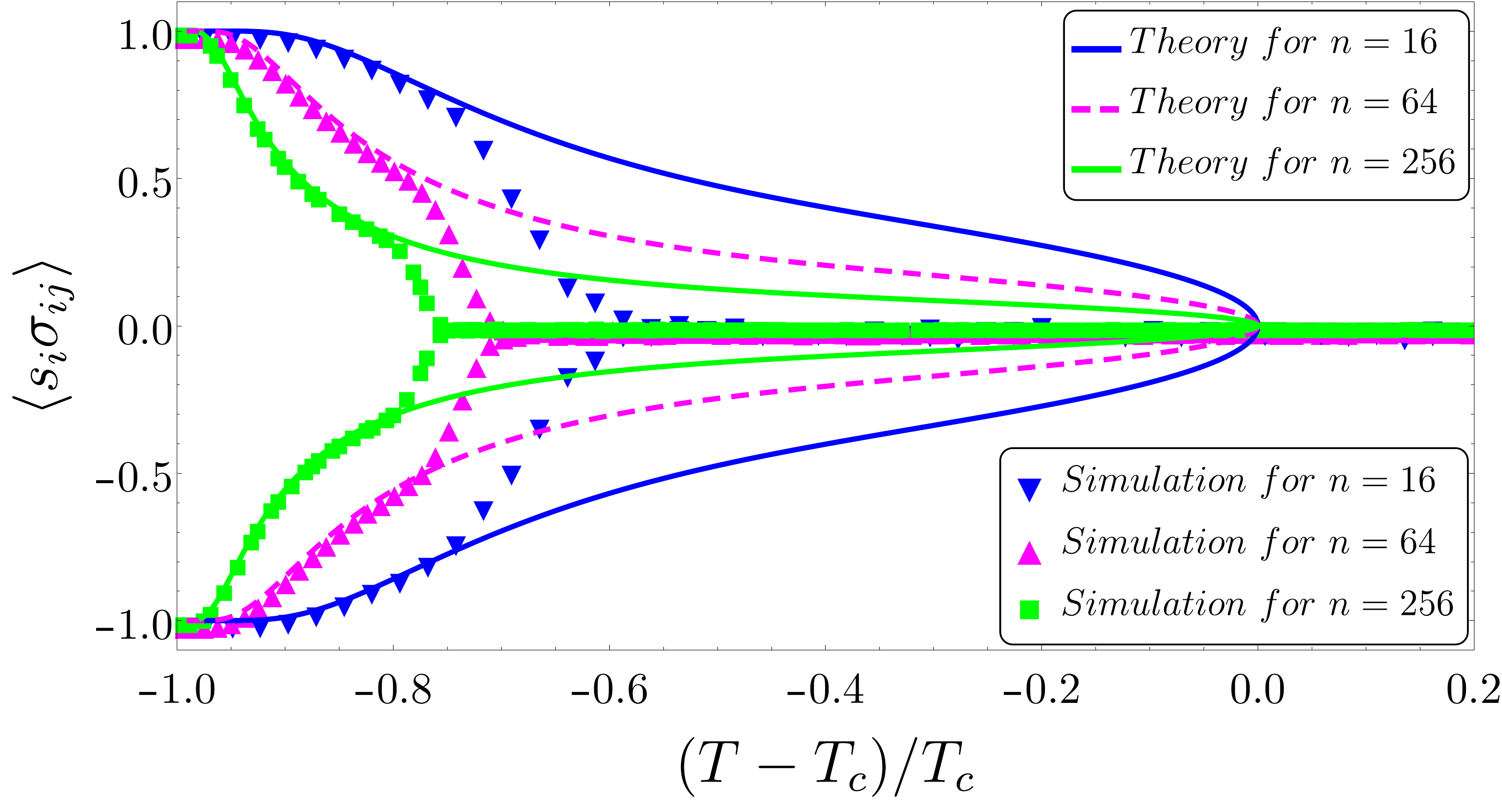}
		\caption{Finite size effect for presented model in  agreement type I and II initial states (Fig.~\ref{fig:Fig1}). The deviation of theory and simulation will be smaller in larger size.   }
		\label{fig:Fig6}
	\end{figure}

	We have checked the finite size effect on the presented model in Fig.~\ref{fig:Fig6}. The deviation of simulation from the theory is very small in low temperatures for all network sizes. This deviation in high temperature is large for small size networks, however when the bigger size is chosen it becomes smaller because the theory line is approaching horizontal axes. In the thermodynamic limit, it will be very small and theory and simulation will be in good agreement. 
	
	\section{CONCLUSIONS}
	Here we present a model that individuals update their opinions and their social relationships to reduce overall social tension (Hamiltonian). By considering social temperature as a measure of the average irrationality of individuals in society, we investigate the effect of this parameter on minimum tension states by mean field approximation. We compared presented model with the thermal Heider balance model \cite{fereshteh} which has been studied in a similar way and the following results have been obtained:
	\begin{itemize}
		\item Phase transition in the presented model is continuous while thermal Heider balance has discrete phase transition.
		
		\item The critical temperature of our model is proportional to the square root of the network’s size, while in thermal Heider balance the critical temperature changes linearly with size \cite{amir}. 
		
		\item Thermal Heider balance has good agreement with simulation in finite sizes but in the coevolutionary balance, this agreement is good for large sizes.
	\end{itemize}
	
	As results show, the differences between coevolutionary and Heider balance models are fundamental. One of the most important differences between these two models is that in Heider balance model the relationship between two people is related to the third person (triangles), while the relationship in coevolutionary balance is only related to the opinions of two people (node-link-node triplet). The number of triangles in the Heider balance model is approximately equal to the size of the network to the power of three ($\binom{n}{3}\approx n^3$), and the number of triplets in the coevolutionary balance model is equal to the number of network links (approximately $\binom{n}{2}\approx n^2$). This difference in the number of triplets indicates that in the Heider model, the number of triplets with a common links is approximately equal to the size of the network. Changing the sign of a relationship has a much greater impact on the overall amount of social tension in the Heider model than in the coevolutionary model [Fig.~\ref{fig:Fig4}]. The effect of average irrationality of individuals (temperature) on the ordered phase is significantly different for these two models. With the increase of temperature, in Heider model, the order disappears abruptly, while in the coevolutionary model this effect slowly destroys the order.

	\begin{acknowledgments}                
		We would like to appreciate M. Saeedian, B. Askari, Z. Moradimanesh and M. Bagherikalhor for constructive discussions. We would like to express our special thanks of gratitude to Center of Excellence in Cognitive Neuropsychology.
		
	\end{acknowledgments}          
	\appendix
	\begin{appendices}
		\section{Deriving Self Consistence Equation from Approximated Partition Function}
		\label{appendix:mean quantity}	
		The main assumption for mean field theory is to neglect fluctuation of microscopic variables around their mean values. We can split the node, edge variables into mean $ \langle s_i\rangle $, $\langle s_i\sigma_{ij}\rangle$ and deviation (fluctuation) $ \delta s_i = s_i - m $, $ \delta (\sigma_{ij}s_j) = \sigma_{ij}s_j - q $ and assumes that the second-order term with respect to the fluctuation $ \delta s_i $, $ \delta (\sigma_{ij}s_j) $  is negligibly small in the interaction energy:
		\begin{equation}
		\begin{aligned}
		\mathcal{H}=&-\sum_{i<j}\left[(\delta s_i+m)(q+\delta(\sigma_{ij}s_j))\right]\\
		&\qquad\qquad-h_1\sum_{i}s_i-h_2\sum_{i<j}\sigma_{ij}s_j,\\
		\approx&n(n-1)mq/2-\Big(h_1+q(n-1)\Big)\sum_{i}s_i\\
		&\qquad\qquad-\Big(h_2+m\Big)\sum_{i<j}\sigma_{ij}s_j.\\
		\end{aligned}
		\end{equation}
		Above is mean field Hamiltonian and the partition function is
		\begin{equation}
		\begin{aligned}
		\mathcal{Z}&=\text{Tr}\,e^{-\beta\mathcal{H}}=e^{\beta n(n-1)mq/2}\Big(2\cosh\Big[\beta (h_1+(n-1)q)\Big]\Big)^n\\
		&\qquad\qquad\qquad\qquad\times \Big(\cosh\left[\beta (h_2+m)\right]\Big)^{n(n-1)/2}.
		\end{aligned}
		\end{equation}
		The Helmholtz free energy is
		\begin{equation}
		\begin{aligned}
		\mathcal{F}=&-k_BT\log\mathcal{Z}\\
		\propto&-k_BTn\log\left\{\cosh\Big[\beta (h_1+ (n-1)q)\Big]\right\}\\
		&-k_BT n(n-1)/2\log\left\{\cosh\left(\beta (h_2+m)\right)\right\}.
		\end{aligned}
		\end{equation}
		We can find self consistence (equation of state) by derivative of free energy with respect to external fields: 
		\begin{equation}
		\begin{aligned}
		m&=\langle s_i\rangle=-\frac{1}{n}\frac{\partial \mathcal{F}}{\partial h_1}\Big|_{h_1=0}=\tanh (\beta  (n-1) q),
		\\
		q&=\langle \sigma_{ij}s_j\rangle=-\frac{2}{n(n-1)}\frac{\partial \mathcal{F}}{\partial h_2}\Big|_{h_2=0}=\tanh (\beta  m).\\
		\end{aligned}
		\end{equation}
		Finally the self consistence equation is
		\begin{equation}
		q=\tanh \Big(\beta \tanh (\beta  (n-1) q) \Big).
		\end{equation}	
	\end{appendices}

\end{document}